\documentclass[prl,twocolumn]{revtex4}

\usepackage{graphicx}

\begin{document}

\title{Absence of the $d$-Density Wave State in 2D Hubbard Model}
\author{
A.\ Macridin$^{1}$,  M. \  Jarrell$^{1}$, and 
Th.\ Maier$^{2}$}

\address{
$^{1}$University of Cincinnati, Cincinnati, Ohio, 45221, USA \\
$^{2}$Oak Ridge National Laboratory, Oak Ridge, Tennessee, 37831, USA }

\date{\today}

\begin{abstract}
  Using the Dynamical Cluster Approximation (DCA) we calculate the
  alternating circulating-current susceptibility and investigate the
  transition to the $d$-density wave (DDW) order in the
  two-dimensional Hubbard model.  The $2 \times 2$ cluster used in the
  DCA calculation is the smallest that can capture $d$-wave order;
  therefore, due to the mean-field character of our calculation, we
  expect to overestimate $d$-wave transition temperatures.  Despite
  this, we found no transition to the DDW state.  In the pseudogap
  region the DDW susceptibility is enhanced, as predicted by the slave
  boson $SU(2)$ theory, but it still is much smaller than the $d$-wave
  pairing susceptibility.

\end{abstract}

\maketitle
\paragraph*{Introduction}

The high $T_c$ cuprates display a variety of unusual properties, which
remain unexplained by conventional theories. The most intriguing
physics occurs at small doping, in the proximity of antiferromagnetism
and superconductivity, and is characterized by non-conventional
behavior of many observables including the spin susceptibility,
optical conductivity, specific heat and transport properties.  Many of
these unusual properties are associated with the presence of a
pseudogap in the one- and two-particle spectra. Photoemission
spectra show that, at small doping and above $T_c$, the
states around the $(\pi, 0)$ point in the Brillouin Zone are gapped
and Fermi segments appear around $(\pi/2, \pi/2)$, suggesting that the
symmetry of the pseudogap in the hole-doped cuprates is consistent
with the $d$-wave symmetry of the superconducting gap~\cite{hpes}.

Based on the cuprate phenomenology, Chakravarty {\em et al.}~\cite{chakra} 
proposed that the pseudogap results from the competition between two ordering 
processes.  One is $d$-wave superconductivity (DSC) and the other is a state 
characterized by long-range order of alternating orbital currents.  The latter 
is a staggered-flux state which breaks the translational and the time-reversal 
symmetry and represents in fact a charge density  with $d$-wave 
symmetry~\cite{nayakddw}, i.e. it is a  $d$-density wave (DDW) state. In this  
scenario the system evolves continuously from the DDW state to the DSC state 
with decreasing temperature or increasing doping, and the two states coexist 
up to optimal doping.  The experimentally observed one-particle spectra in the 
pseudogap region can be well understood on the basis of the DDW 
state~\cite{chakraarpes}, which makes it a very appealing candidate for the 
origin of the pseudogap physics.  Other properties of the cuprates, such as 
the resonant peak in the superconducting state and the doping dependence of 
superfluid density seem also to be well captured by this model~\cite{chakraass}. 
Recently it was  proposed the interplay of the DDW and the inter-planar 
tunneling of Cooper pairs to be responsible for the $T_{c}$ dependence on the 
number of $CuO_2$ layers which characterizes different 
materials ~\cite{chakrtunel}.  In principle, the presence of the DDW state has 
subtle experimental consequences, such as the formation of a magnetic moment 
associated with the orbital currents~\cite{chakraorb}. The interpretation of 
the experimental data in this respect however is still controversial.

Whereas the theory of the DDW is phenomenological, slave boson theory
holds the promise of a microscopic basis which may explicitly consider static 
or fluctuating DDW order, as well as $d$-wave pairing. These are uncontrolled 
theories for the $t-J$ model, which is equivalent to the strong coupling limit 
of the  Hubbard Hamiltonian.  They are influenced by P.W. Anderson's 
{\em et al.}~\cite{rvb} idea of resonance valence bond state.  The charge 
and the spin degrees of freedom are separated by introducing  auxiliary slave 
bosons.  The resulting mean-field theories explicitly decouple the fermion 
hopping along the bonds, the fermion pairing  and the bosonic field, and 
produce phase diagrams similar to the experimental one.

We briefly discuss the main results of slave boson theories.  The $t-J$ 
Hamiltonian has a local $SU(2)$ symmetry at half filling~\cite{andersonsu2}.  
As a result of this symmetry the $\pi$-flux state (a staggered-flux state
with the flux per plaquette equal to $\pi$) and the $d$-wave paring
state are degenerate in the undoped model. Doping breaks the symmetry
to $U(1)$ and the $d$-wave state becomes energetically favored~\cite{kotliar}.  The
$d$-wave state is characterized by a finite value of the fermion pairing 
operator and a real superconductor emerges below the condensation temperature 
of the bosons.  At low doping ($\delta \leq 0.05$) and for $T>0$ the 
$d$-wave pairing becomes unstable towards the $\pi$-flux or staggered-flux 
state of spinons~\cite{ubbenslee}.  This is the standard picture of the 
$U(1)$ slave boson mean field theory of the $t-J$ model.  However, the 
inclusion of other terms in the mean-field decoupling, such as the holon's 
flux~\cite{hamada}, results in a existence of a DDW state at finite doping 
and above $T_c$, but exclude the coexistence of the two states.

One of the drawbacks of the $U(1)$ theory is that its solution is not
stable against the fluctuations of the gauge field~\cite{ubbenslee2}. 
Fluctuations are especially important at small doping, where the energy 
difference between states connected via a $SU(2)$ transformation is very 
small (since they are degenerate at zero doping).  Therefore all these
states have an important contribution in the determination of the free
energy.

P.A.\ Lee {\em et al.} developed a slave boson mean field theory which
is $SU(2)$ symmetric at finite doping~\cite{leesu2}. The price paid is 
that one must deal with two slave bosonic fields and three constraints. 
The advantage of this approach is that the $SU(2)$ mean-field solution is
likely superior at small doping, since it accounts better for the
fluctuations between different low energy $SU(2)$ connected states.
Their solution for the pseudogap region is a staggered-flux of
fermions which is gauge equivalent with the $d$-wave pairing of
fermions.  However, the fermion staggered-flux state is not the same
as the staggered-flux state of electrons (or the DDW state), and
neither breaks time-reversal nor translation symmetry.  Therefore, in
$SU(2)$ theory, the pseudogap is not a broken symmetry state with long
range order as it is DDW, but is rather is characterized by strong
spatial and dynamic fluctuations between $d$-wave, $s$-flux and other
$SU(2)$ related states.

The goal of this paper is to investigate the interplay between DDW and
DSC order in the 2D Hubbard model.  Using the Dynamical Cluster
Approximation (DCA)\cite{m_hettler_98,m_hettler_00} we calculate the
response functions associated with these two types of order.  The DCA
systematically adds non-local corrections to the Dynamical Mean Field
Approximation (DMFA)\cite{t_pruschke_95,a_georges_96} by mapping the
lattice onto a finite-size periodic cluster.  The DCA mapping from the
lattice to the cluster is accomplished by coarse-graining all the
internal propagators in irreducible Feynman graphs in reciprocal
space.  Correlations at short length scales, within the cluster, are
treated explicitly with a quantum Monte Carlo (QMC) simulation, while
those at longer length scale are treated at the mean field level.  Due
to the residual mean field character of our approximation, we expect
our calculation to {\em{overestimate the transition temperatures of
both the DDW and DSC critical temperatures}}.

Generally, we expect to see the the most pronounced mean-field behavior 
from the smallest cluster that can reflect the broken symmetry.  A similar 
situation occurs in DMFA simulations of the antiferromagnetic phase of 
the Hubbard model, where N\'eel order is possible since the impurity spin 
and the mean-field host may have opposite spin orientations.  Since 
non-local fluctuations are suppressed, the DMFA overestimates the N\'eel 
transition temperature.  In the present case, the $N_c=4$ (i.e.\ 
$2 \times 2$) is the smallest possible cluster allowing for $d$-wave 
pairing or a circulating current.  Orbital antiferromagnetism is possible 
since the moment in the cluster and the host can have opposite orientations.  
Since fluctuations on longer length scales are suppressed, we would expect 
to overestimate both the $d$-wave superconducting and the $d$-density wave 
transition temperatures. 

\paragraph*{Formalism} We present  DCA calculations for the conventional 
2D Hubbard model describing the dynamics of electrons on a square lattice. 
The model
\begin{equation}
\label{eq:ham}
H= -t \sum_{<i,j>, \sigma} c^{\dagger}_{i,\sigma}c_{j,\sigma}+U \sum_i
 n_{i\downarrow} n_{i \uparrow}~,
\end{equation} 
is characterized by a hopping integral $t$ between nearest neighbor
sites and a Coulomb repulsion $U$ two electrons feel when residing on
the same site. As the energy scale we set $t=0.25{\rm eV}$ so that the
band-width $W=8t=2{\rm eV}$, and study the intermediate coupling regime 
$U=W$.  We study the dynamics on short length-scales by setting the 
cluster size to $N_c=4$. This cluster size is large enough to capture the 
qualitative low-energy physics of the cuprate 
superconductors \cite{t_maier_00b, m_jarrell_01b}.  The corresponding
phase diagram resembles the generic phase diagram of 
cuprates~\cite{ m_jarrell_01b}, displaying regions characterized by
antiferromagnetism, $d$-wave superconductivity, Fermi liquid and
pseudogap regimes, in qualitative agreement with experimental results.

In this paper we calculate the static (i.e. $\omega=0$) susceptibilities 
which correspond to the circulating current (cc) operator,
\begin{equation}
\label{eq:cc}
W= i\sum_{k,\sigma} g(k)  c^{\dagger}_{k+Q,\sigma}c_{k,\sigma}~,
\end{equation} 
\noindent and respectively to the $d$-wave pairing operator,
\begin{equation}
\label{eq:dsc}
P= \sum_{k,\sigma} g(k)  c_{k\downarrow}c_{-k \uparrow}~,
\end{equation} 
where $g(k)=\cos(k_x)-\cos(k_y)$ is the $d$-wave symmetry factor.

\paragraph*{Results}

\begin{figure}[t]
  \centerline{ \includegraphics*[width=3.3in]{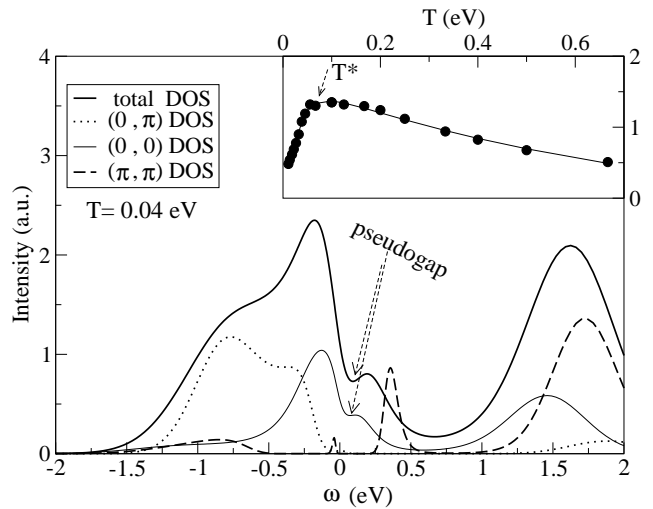}}
\caption{The one particle total and K dependent DOS at $\delta=0.05$ doping.
  Inset: The uniform magnetic susceptibility versus $T$. The maximum
  defines the pseudogap temperature $T^*$.}
\label{fig:doscc}
\end{figure}

The pseudogap temperature, $T^*$ is determined from the maximum in the
uniform magnetic susceptibility (see the inset in Fig.~\ref{fig:doscc}) 
when accompanied by a suppression of spectral weight in the DOS.  We show 
this in Fig.~\ref{fig:doscc} where the total and the $K$-dependent DOS, 
below $T^*$, at $\delta=0.05$ doping is plotted.  The DCA on a $N_c=4$ 
cluster implies a coarse graining of the Brillouin Zone in four cells 
around $K=(0,0)$, $(0,\pi)$, $(\pi,0)$ and $(\pi,\pi)$ and the $K$-dependent 
DOS corresponds to the average over all $k$ belonging to a coarse-grained 
cell of the single particle spectra $A(k,\omega)$.  This poor resolution 
in the reciprocal space allows to study only  the gross features in the 
single-particle spectra. Despite this, it can be seen from Fig.~\ref{fig:doscc} 
that the pseudogap in the total DOS is a result of the suppression of spectral 
weight in the cell at $(0,\pi)$.  Therefore, we believe that our calculations 
capture well the experimentally observed features of the pseudogap.

\begin{figure}[t]
\centerline{
\includegraphics*[width=3.3in]{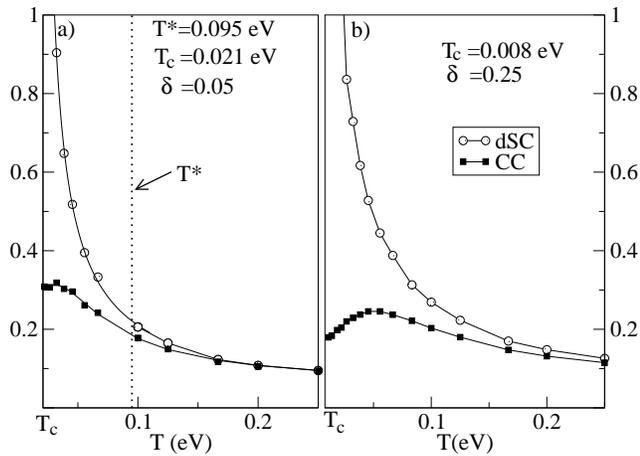}}
\caption{The $d$-wave pairing (circles) and the circulating current
(squares) susceptibilities versus temperature, at a) $5\%$ and b) $25\%$ doping.
The dotted vertical line in a) is at the pseudogap temperature $T^*$.} 
\label{fig:cc-sc}
\end{figure}

However, our calculations show that these features are not a
consequence of the DDW state.  In Fig.~\ref{fig:cc-sc}-a we plot both
the $d$-wave pairing susceptibility and the cc-susceptibility versus
temperature, at $\delta=0.05$ doping. The pairing susceptibility
diverges at $T_c$, indicating a $d$-wave superconducting
instability.  The $cc$-susceptibility does not diverge, indicating
the absence of a possible transition to the DDW state.

At large temperatures the $d$-wave pairing and the cc-susceptibilities
are degenerate, and they both increase with decreasing temperature.
In the pseudogap region (left side of dotted line) the $d$-wave pair
field susceptibility is much larger than the cc-susceptibility.  Close
to $T_c$, the cc-susceptibility saturates and starts even decreasing
with decreasing temperature.  The fact that in the pseudogap region
both $d$-wave and the cc-susceptibilities are enhanced show that
fluctuations between these states are significant, as it was predicted
by the $SU(2)$ theory~\cite{leesu2}.

In Fig.~\ref{fig:cc-sc}-b we show the $d$-wave and the cc-susceptibilities 
at $\delta=0.25$ doping. 
No pseudogap~\cite{m_jarrell_01b} is observed at this doping.
We notice that, starting well above $T_c$, the cc-susceptibility decreases 
with decreasing temperature. This behavior is different from the one observed 
at small doping where the cc-susceptibility increases with decreasing $T$ 
up to $T_c$. We therefore conclude that in the overdoped region the 
fluctuations between DSC and DDW above $T_c$ are much less important.

Calculations (not shown here) with other values of the parameters
(different values of $U$), or with the inclusion of next-nearest-neighbor 
hopping corresponding to both electron and hole doping, exhibit similar 
results.  A divergent DDW susceptibility is never found. Our results are 
consistent with renormalization group studies \cite{honerkamp} where no 
divergence of the cc-susceptibility is found, and with a mean-field 
treatment of an extended Hubbard model\cite{nayak} where an additional 
correlated hopping term was necessary to stabilize the DDW state.

\paragraph*{Conclusions}
We present a DCA calculation of the two-dimensional Hubbard model, focusing 
on the competition between DDW and DSC orders. We showed previously that the DCA 
calculation captures the generic features of the pseudogap region as seen 
in the photoemission and magnetic measurements. Nevertheless, as we show here,
these properties are not a consequence of the existence of DDW state.  The 
DCA should overestimate any DDW transition temperature but, despite this, 
we found no  transition to such state. 

We also found that both the $cc$-susceptibility and the $d$-wave pairing 
susceptibility are enhanced in the pseudogap region, indicating that the 
fluctuations between these states is significant. This is not true in  the
overdoped region, where we found that the $cc$-susceptibility is suppressed 
above $T_c$.

\paragraph*{Acknowledgment}

We acknowledge useful conversations with G.~Baskaran, C.~Honerkamp,
P.~Lee, and W.~Putikka.  This research was supported by the NSF grant
DMR-0312680.  Part of this research was performed by TM as Eugene P.
Wigner Fellow and staff member at the Oak Ridge National Laboratory,
managed by UT-Battelle, LLC, for the U.S. Department of Energy under
Contract DE-AC05-00OR22725.

\end{document}